\documentclass{article}
\usepackage{ijcai25}

\usepackage{times}
\usepackage{soul}
\usepackage{url}
\usepackage[hidelinks]{hyperref}
\usepackage[utf8]{inputenc}
\usepackage[small]{caption}
\usepackage{graphicx}
\usepackage{amsmath}
\usepackage{amsthm}
\usepackage{booktabs}
\usepackage{algorithm}
\usepackage{algorithmic}
\usepackage[switch]{lineno}

\usepackage{amsfonts}
\usepackage{multirow}
\usepackage{pifont}

\title{SourceDetMamba: A Graph-aware State Space Model for Source Detection in Sequential Hypergraphs}

\author{
Le Cheng$^{1,2}$\and
Peican Zhu$^1$\thanks{Corresponding authors.}\and
Yangming Guo$^3$\and
Chao Gao$^1$\and
Zhen Wang$^{3}$\footnotemark[1]\And
Keke Tang$^4$\footnotemark[1]\\
\affiliations
$^1$School of Artificial Intelligence, Optics and Electronics, Northwestern Polytechnical University\\
$^2$School of Computer Science, Northwestern Polytechnical University\\
$^3$School of Cybersecurity, Northwestern Polytechnical University\\
$^4$Cyberspace Institute of Advanced Technology, Guangzhou University\\
\emails
\{ericcan, w-zhen\}@nwpu.edu.cn,
tangbohutbh@gmail.com
}

\begin{document}

\maketitle

\begin{abstract}
Source detection on graphs has demonstrated high efficacy in identifying rumor origins. Despite advances in machine learning-based methods, many fail to capture intrinsic dynamics of rumor propagation. In this work, we present SourceDetMamba: A Graph-aware State Space Model for Source Detection in Sequential Hypergraphs, which harnesses the recent success of the state space model Mamba, known for its superior global modeling capabilities and computational efficiency, to address this challenge. Specifically, we first employ hypergraphs to model high-order interactions within social networks. Subsequently, temporal network snapshots generated during the propagation process are sequentially fed in reverse order into Mamba to infer underlying propagation dynamics. Finally, to empower the sequential model to effectively capture propagation patterns while integrating structural information, we propose a novel graph-aware state update mechanism, wherein the state of each node is propagated and refined by both temporal dependencies and topological context. Extensive evaluations on eight datasets demonstrate that SourceDetMamba consistently outperforms state-of-the-art approaches.
\end{abstract}

\section{Introduction}
Source nodes detection on graphs provides a feasible approach to addressing societal challenges, such as rumor source identification, while simultaneously posing fundamental mathematical problems  \cite{shah2011rumors,zhu2022locating}. To address this issue, early methods such as LPSI \cite{wang2017multiple}, OJC \cite{zhu2017catch}, and MLE \cite{pinto2012locating} utilize source centrality theory \cite{shah2011rumors} and maximum likelihood estimation \cite{yang2020locating} to identify propagation sources. More recently, with the advancement of machine learning, particularly GCNs \cite{kipf2016semi}, researchers embed user information and network topology into optimized models \cite{dong2019multiple,ling2022source,wang2022invertible}, continuously achieving state-of-the-art performance across various social network scenarios \cite{bao2024graph,feng2024backdoor}.

\begin{figure}[t]
  \centering
  \includegraphics[width=\linewidth]{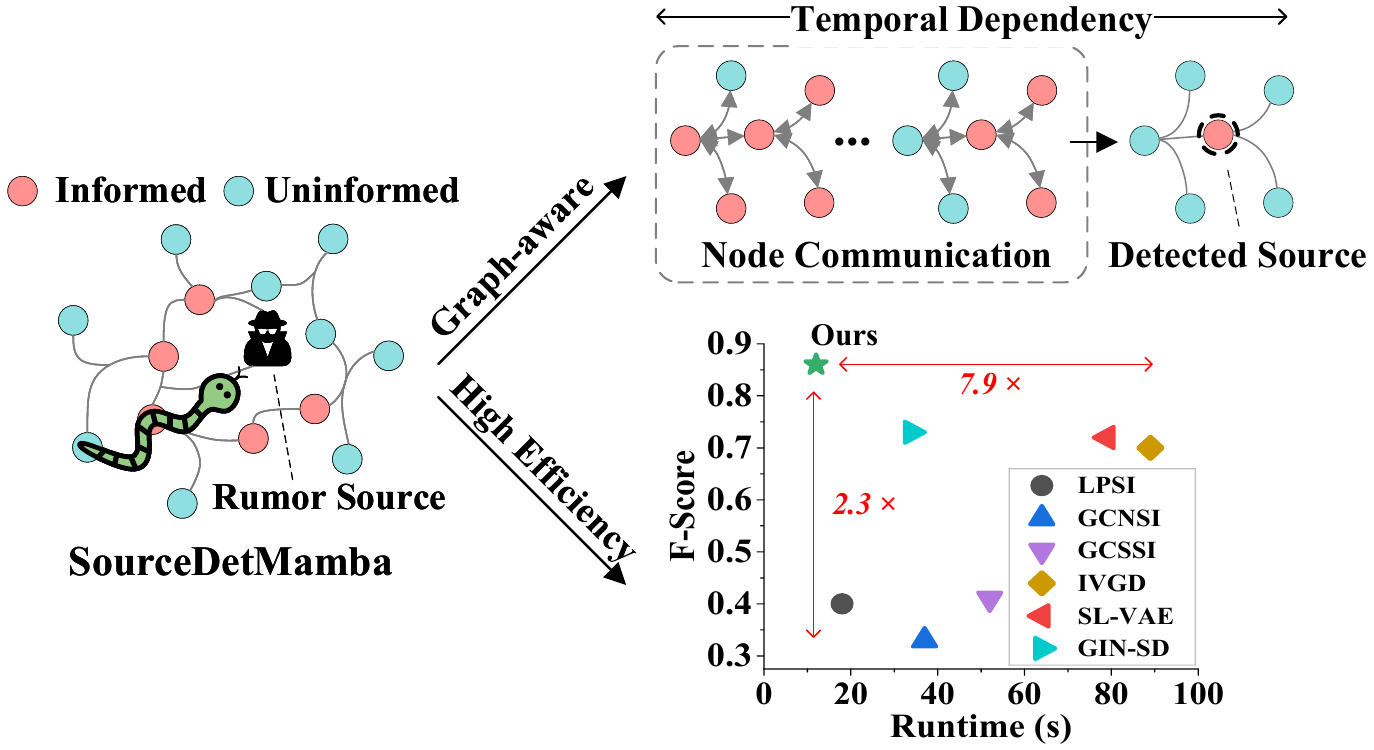}
  \caption{
We propose a Mamba-based solution
for source detection in sequential hypergraphs, i.e., SourceDetMamba, which demonstrates significant improvements in both accuracy  and  efficiency compared to baseline methods.
  }
  \label{mamba_example}
  \vspace{-1mm}
\end{figure}

However, most of current solutions are based on a single network snapshot. These approaches either heavily rely on handcrafted features for node classification without learning the propagation patterns and trends; or first fit a propagation model to the snapshot before performing source detection. Unfortunately, the process of fitting the propagation model introduces extra errors, hence limiting the subsequent source identification performance. This raises an intuitive question: \textit{can we directly model propagation patterns in reverse order from sequential network snapshots to detect sources}?

Addressing this problem introduces new challenges. First, existing sequence models struggle with this task. Traditional models like RNNs and LSTMs suffer from limited performance due to their inability to handle long-term dependencies effectively \cite{liang2024pointmamba}. While the Transformer architecture improves performance by leveraging attention mechanisms to compute weights for each point in the sequence, its quadratic complexity results in high computational costs for large-scale scenarios \cite{behrouz2024graph}. Second, directly feeding the features of all nodes in a single network snapshot as a sequence element into the sequential model results in the loss of critical topological information.

We note that state space models (SSMs) have recently made remarkable progress. For instance, the structured state space model (S4) enhances the ability to handle long-range dependencies, achieving powerful representation capabilities with only linear complexity \cite{gu2022efficiently,gu2021combining}. Building on S4, another pioneer that has garnered significant attention, Mamba, a selective SSM, introduces time-varying parameters to enable content-aware reasoning \cite{gu2023mamba}. Moreover, Mamba employs a hardware-aware algorithm that allows parallel computation, ensuring efficient training and inference. It has demonstrated impressive performance, matching transformer-level results while maintaining linear computational complexity. However, directly applying Mamba to source detection via sequential snapshots yields suboptimal results. The primary challenge is how to ensure that Mamba effectively incorporates topological information from each network snapshot while learning propagation patterns. Addressing this issue is critical before deploying Mamba for effective source detection.

In this paper, we propose SourceDetMamba: a graph-aware state space model for source detection in sequential hypergraphs. First, hypergraphs are utilized to model the higher-order interactions in real-world networks, such as group chats on social media \cite{yin2022dynamic,jiao2024enhancing}. Next, to directly learn the rumor propagation patterns, multiple sequential network snapshots generated during the rumor propagation process are fed into Mamba in a reverse chronological order to capture the underlying dynamics. Finally, to ensure that both the sequential information and the topological structure of each snapshot are effectively incorporated, we innovatively introduce a graph-aware state update mechanism, where each node's state is determined by both the sequential input and its neighborhood. We evaluate our approach on eight public datasets. Extensive experiments demonstrate that our method outperforms SOTA approaches, establishing new benchmarks in source detection tasks.

Overall, our contributions are summarized as follows:
\begin{itemize}
    \item We design a unified framework, SourceDetMamba, which leverages the recent advances in state space models to directly learn propagation patterns from reverse-ordered network snapshots.
    \item We propose a novel graph-aware state update mechanism that integrates sequential propagation dynamics with the underlying network topology, enabling state space models to be effectively applied to sequential tasks on graphs.
    \item We show by comprehensive experiments on eight real-world datasets that SourceDetMamba consistently outperforms state-of-the-art methods in terms of both accuracy and efficiency.
\end{itemize}

\section{Related Work}
\subsection{Snapshot-based Source Detection}
Network snapshots have become the preferred foundation for source detection methods in recent years due to the convenience and capability to capture user states and network topology \cite{cheng2024gin}. Based on source centrality theory \cite{prakash2012spotting,shah2011rumors}, LPSI identifies locally prominent nodes via label propagation \cite{wang2017multiple}, EPA iteratively infers infection times \cite{ali2019epa}, and OJC optimizes Jordan centrality \cite{zhu2017catch}. Although computationally cheap, these heuristics struggle with the complex attributes and diverse dynamics of real social networks \cite{cheng2024heuristic}. Machine learning offers a promising alternative \cite{pei2024multi,pei2020active}. Utilizing GCNs, GCNSI \cite{dong2019multiple} and SIGN \cite{li2021propagation} incorporate LPSI as a preprocessing step and employ user states for node classification, while GCSSI targets the wavefront of infection \cite{dong2022wavefront}. From a model architecture perspective, ResGCN adds residual links to deepen message passing \cite{shah2020finding}. Nevertheless, these approaches still under-learn intrinsic diffusion patterns. To address this limitation, IVGD \cite{wang2022invertible} and SL-VAE \cite{ling2022source} model heterogeneity via graph diffusion. However, the extra error in fitting dynamics limits accuracy, and heavy attention modules further raise computation.


\subsection{State Space Models}


Recently, state space models (SSMs) have emerged as a promising paradigm for sequence modeling, offering an efficient alternative to CNNs and Transformers \cite{gu2021combining}. For instance, Structured State Space for Sequences (S4) leverages the HIPPO matrix to construct a state representation that effectively captures recent inputs while gradually decaying older information \cite{gu2020hippo}. This design allows S4 to process long sequences with linear complexity. Building on S4, Mamba introduces adaptive parameters, enabling dynamic filtering of relevant and irrelevant information. Additionally, its hardware-aware design supports efficient storage and computation through parallel scanning, achieving Transformer-level performance while maintaining linear complexity \cite{gu2023mamba}. In graph-related tasks, Graph Mamba extends Mamba framework to single-graph by incorporating neighborhood tokenization, token ordering, and a selective SSM encoder \cite{behrouz2024graph}. However, the exploration of state space models on sequential graphs remains an open challenge, leaving a gap in methods designed for such tasks.

\begin{figure*}[!htbp]
  \centering
  \includegraphics[width=0.96\linewidth]{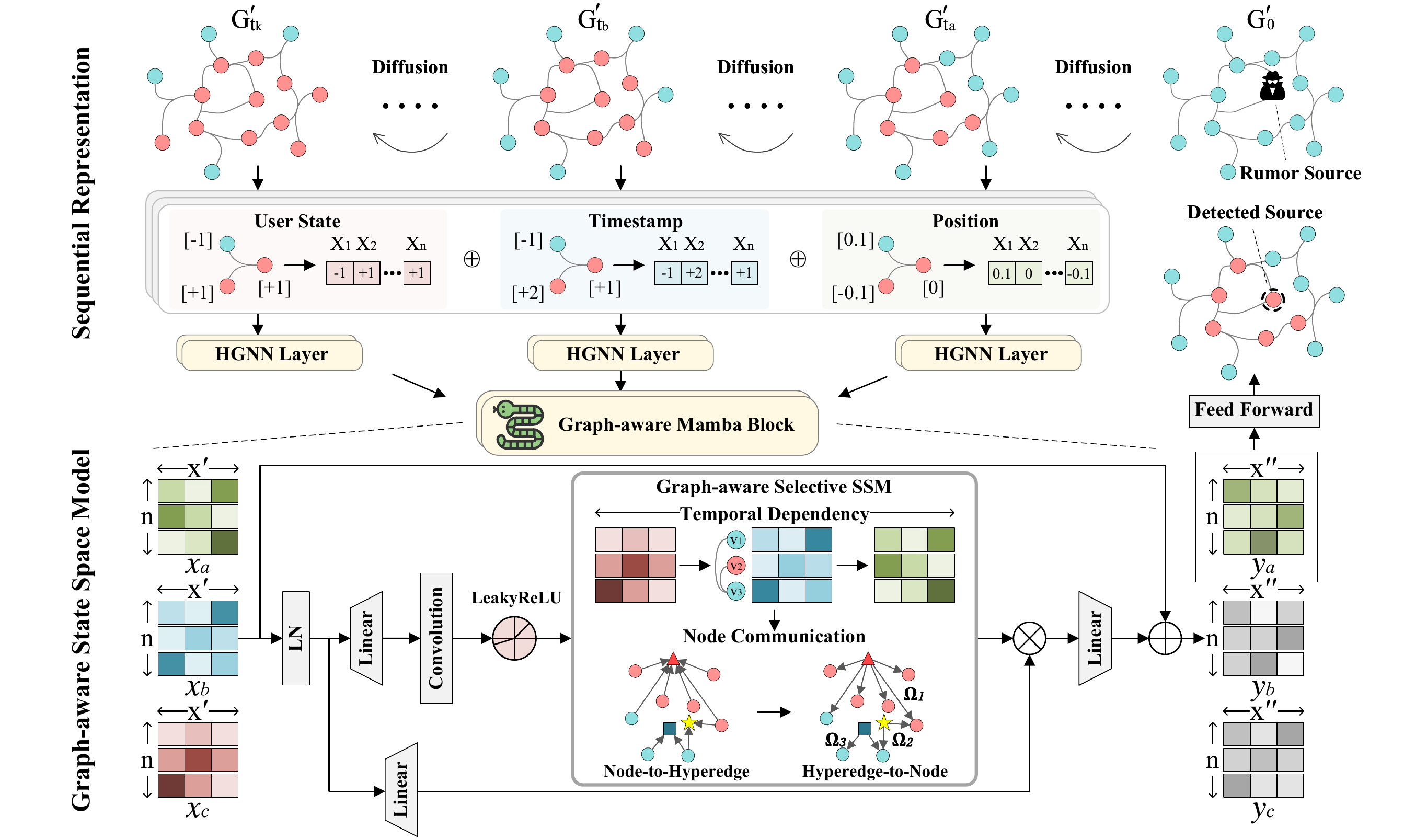}
  \caption{Framework of SourceDetMamba. Starting with hypergraph propagation snapshots, node features are embedded and refined via hypergraph neural networks to capture high-order interactions, which are then input into the graph-aware state space model in a reverse order to learn propagation patterns. Finally, the source nodes are identified based on the last element of the output sequence.}
  \label{framework}
  \vspace{-2mm}
\end{figure*}

In this work, we focus on proposing a state space model for source detection via sequential hypergraph snapshots, which effectively captures propagation patterns through incorporating the topological information of each snapshot.

\section{Problem Formulation}
\paragraph{Preliminary on Hypergraphs.}
Hypergraphs extend the pairwise graph structure by accommodating higher-order interactions through hyperedges, enabling the representation of complex relationships in social networks. Formally, a social network can be abstracted as a hypergraph $G$ = $(V,E,\mathbf{\Omega})$, where $V$ = $\{v_1, v_2, ...,v_n\}$ denotes the set of nodes, and $E$ = $\{e_1,e_2,...,e_m\}$ represents the set of hyperedges, with each hyperedge $e_i$ = $\{v_{i1},v_{i2},...,v_{ij}\}$ consists of a subset of nodes, where $i \in \{1,2,...,m\}$ and $j \in \{1,2,...,n\}$. $\mathbf{\Omega} \in \mathbb{R}^{m \times m}$ is a diagonal matrix, where each element corresponds to the weight of a hyperedge. Given the relationship between nodes and hyperedges, the hypergraph $G$ can be represented by an incidence matrix $\textit{\textbf{H}} \in \mathbb{R}^{n \times m}$, each entry $\textbf{\textit{H}}_{ve}$ indicates the participation of a node $v$ in a hyperedge $e$:
\begin{align}
\textbf{\textit{H}}_{ve} & = \left\{\begin{array}{ll}
1, & v \in e \\
0, & \text {otherwise}.
\end{array}\right.
\end{align}

Consequently, the degree of nodes and hyperedges are expressed using diagonal matrices $\textit{\textbf{D}}_V \in \mathbb{R}^{n \times n}$ and $\textit{\textbf{D}}_E \in \mathbb{R}^{m \times m}$, where each entry $\textit{\textbf{D}}_{vv}$ = $\sum_{e=1}^m \textit{\textbf{H}}_{ve}$ and $\textit{\textbf{D}}_{ee}$ = $\sum_{v=1}^n \textit{\textbf{H}}_{ve}$, respectively.

\paragraph{Propagation Process on Hypergraphs.}
The rumor propagation process in social networks evolves over time $t$, where the initially uninformed source nodes transition to the informed state at $t$ = 0, triggering the spread of information. Unlike the pairwise interactions modeled in conventional graphs, propagation on hypergraphs encompasses both low-order interactions between individual nodes and high-order interactions within hyperedges, representing peer influence or collective pressure among groups. During the low-order interaction phase, an informed node $v_i$ disseminates information to its neighbors with a probability $p_i$. In the high-order interaction phase, a node transitions from the uninformed to the informed state if the cumulative influence from its neighbors exceeds a threshold $p_\triangle$. Several classical models, such as SI, SIR, IC, and LT \cite{battiston2020networks,de2020social}, have been proposed to simulate the propagation process on hypergraphs.

In summary, the propagation on hypergraphs can be represented as a time series $\{G'(t),t \geq 0\}$, where $G'(t)$ is the network snapshot at time $t$, encapsulating information such as network structure, node states, and propagation information.

\paragraph{Source Detection in Sequential Hypergraphs.}
As the propagation evolves, a series of network snapshots are captured at distinct time points $T$ = $\{t_a, t_b,\cdots, t_k\}$, which are not necessarily continuous or evenly spaced. Consequently, source detection on sequential hypergraphs can be defined as:
\begin{equation}
\hat{s}=f(\{G'(t_i)\}_{t_i \in T}),
\end{equation}
where $f(\cdot)$ denotes the source detection method, and $\hat{s}$ represents the identified source set.


\paragraph{Main Challenges of Rumor Source Detection in Sequential Hypergraphs.}


\begin{itemize}
    \item[\textbf{(1)}] \textbf{Modeling Sequential Hypergraphs and Node Feature Embedding.} Effectively modeling sequential hypergraphs while embedding node features to capture high-order interactions remains a formidable challenge.
    \item[\textbf{(2)}] \textbf{Integration of Structural Properties into Node Feature Updates.} Although Mamba has demonstrated strong sequential modeling capabilities, directly feeding node features from a snapshot as a sequence element results in the loss of critical topological information.
    
\end{itemize}

\section{Method}
We propose SourceDetMamba, a framework explicitly designed to address the challenges above. Specifically, we introduce the \textbf{Sequential Representation and Feature Fusion} module for Challenge (1), and design a \textbf{Graph-aware State Space Model}  to handle Challenge (2).


\subsection{Sequential Representation and Feature Fusion}
Throughout the propagation process, network snapshots are captured at discrete time points $T$ = $\{t_a, t_b,\cdots, t_k\}$. Within each snapshot, nodes are classified into two subgraphs: $G^+_{t_i}$ (Informed) and $G^-_{t_i}$ (Uninformed). To maximize the utilization of essential information and reinforce comparative features, we embed key attributes, including user states, diffusion timestamps, and positional information, into the initial feature representations.

\paragraph{User State Information $\textbf{\textit{X}}_i^1$.} 
The state of each node is defined by its participation in the rumor propagation process. Specifically, node $v_i \in G^+_{t_i}$ if it has informed the rumor; otherwise, $v_i \in G^-_{t_i}$. Accordingly, $\textbf{\textit{X}}_i^1$ is expressed as:  
\begin{equation}
\textbf{\textit{X}}_i^1 = \left\{\begin{array}{ll}
+1, & v_i \in G^+_{t_i} \\
-1, & \text {otherwise}.
\end{array}\right.
\end{equation}

\paragraph{Diffusion Information $\textbf{\textit{X}}_i^2$.} Social platforms like Facebook and Twitter attach timestamps to user posts, providing critical information for source detection. Hence, diffusion information $\textbf{\textit{X}}_i^2$ is defined as:
\begin{equation}
\textbf{\textit{X}}_i^2 = \left\{\begin{array}{ll}
\ \ t_{v_i}, & v_i \in G^+_{t_i} \\
-1, & \text {otherwise}.
\end{array}\right.
\end{equation}

\paragraph{Positional Information $\textbf{\textit{X}}_i^3$.}
Existing GCN models learn structural node information with invariant node positions \cite{srinivasan2019equivalence}. However, relative positions between nodes play a critical role in global message propagation. Leveraging the generative capacity of Laplacian positional encodings (PEs) \cite{dwivedi2020benchmarking}, we incorporate it as the positional feature of nodes. Inspired by source centrality theory, we focus on calculating PEs for the infected subgraph $G^+_{t_i}$ rather than the entire graph, highlighting the positional significance of sources. The symmetric normalized Laplacian matrix for $G^+_{t_i}$ is defined as:
\begin{equation}
\textit{\textbf{L}}^{sym}_+ = \textit{\textbf{I}} - \textit{\textbf{D}}^{-\frac{1}{2}}_{V+} \textit{\textbf{H}}_+ \textit{\textbf{D}}^{-1}_{E+} \textit{\textbf{H}}_+^T \textit{\textbf{D}}^{-\frac{1}{2}}_{V+},
\end{equation}  
where \( \textit{\textbf{I}} \) denotes the identity matrix, \( \textit{\textbf{H}}_+ \) is the incidence matrix of the subgraph \( G^+_{t_i} \), \( \textit{\textbf{D}}_{V+} \) and \( \textit{\textbf{D}}_{E+} \) represent the degree matrices of nodes and hyperedges, respectively. The Laplacian matrix \( \textit{\textbf{L}}^{sym}_+ \) can further be decomposed as:  
\begin{equation}
\bigtriangleup_{\textit{\textbf{L}}_+^{sym}} = \mathbf{\Gamma} \mathbf{\Lambda} \mathbf{\Gamma}^T, 
\end{equation}
where $\mathbf{\Gamma}$ consists of eigenvectors and $\mathbf{\Lambda}$ is the diagonal matrix of eigenvalues. The eigenvectors associated with the smallest non-zero eigenvalues provide the positional encodings, which we adopt as the node-level positional feature \( \textbf{\textit{X}}_i^3 \):
\begin{equation}
\textbf{\textit{X}}_i^3 = \left\{\begin{array}{ll}
\mathbf{\Gamma}_i, & v_i \in G^+_{t_i} \\
-1, & \text {otherwise}.
\end{array}\right.
\label{X_3}
\end{equation}

Finally, the initial feature of node \( v_i \) is obtained by concatenating the individual components:  
\begin{equation}
\textbf{\textit{X}}_i = \left[\|_{x=1}^3 \textbf{\textit{X}}_i^x\right].
\end{equation}

\paragraph{Sequential Feature Preprocessing.}
For the derived sequential hypergraph features $\{\textbf{\textit{X}}(t_i)\}_{t_i \in T}$. Unlike the feature aggregation process in pairwise graphs, hypergraphs require a unique approach due to the role of hyperedges as mediums for interactions among multiple nodes. To address this, we employ the Hypergraph Neural Network (HGNN)  \cite{feng2019hypergraph,gao2022hgnn+,bai2021hypergraph} that aggregates features in two stages: node-to-hyperedge and hyperedge-to-node. The process is formulated as:
\begin{equation}
    \textbf{\textit{X}}^{(l+1)} = \sigma \left( \textit{\textbf{D}}_V^{-1/2} \textit{\textbf{H}} \mathbf{\Omega}\textit{\textbf{D}}_E^{-1} \textit{\textbf{H}}^\top \textit{\textbf{D}}_V^{-1/2} \textbf{\textit{X}}^{(l)} \textbf{\textit{W}}^{(l)} \right),
    \label{hgnn}
\end{equation}
where \( \textbf{\textit{X}}^{(l)} \) represents the node features at layer \( l \), \( \textit{\textbf{H}} \) is the incidence matrix, \( \textit{\textbf{D}}_V \) and \( \textit{\textbf{D}}_E \) are the diagonal degree matrices of nodes and hyperedges, respectively. \( \mathbf{\Omega} \) denotes the hyperedge weight matrix, and \( \textbf{\textit{W}}^{(l)} \) is the trainable parameters. The activation function \( \sigma(\cdot) \) introduces non-linearity, enabling the model to capture the intricate high-order interactions effectively. Through Eq. (\ref{hgnn}), the initial feature sequence $\{\textbf{\textit{X}}(t_i)^{n \times x}\}_{t_i \in T}$ is projected into $\{\textbf{\textit{X}}'(t_i)^{n \times x'}\}_{t_i \in T}$, serving as the foundation for downstream tasks.

\subsection{Graph-aware State Space Model}
\paragraph{State Space Model.}
By capturing sequential dynamics through explicitly maintaining a latent state that evolves over time, State Space Models (SSMs) map an input sequence $x_t \in \mathbb{R}$ to a latent state representation $h_t \in \mathbb{R}^N$, and subsequently derive a predicted output sequence $y_t \in \mathbb{R}$. While continuous SSMs provide a mathematically elegant framework, finding an analytical solution for the state representation $h_t$ remains a significant challenge. Moreover, the majority of practical sequences, such as textual data, are inherently discrete. Consequently, discrete SSMs are adopted with four parameters ($\Delta$, \textbf{\textit{A}}, \textbf{\textit{B}}, \textbf{\textit{C}}), the sequence-to-sequence transformation can be expressed as:
\begin{equation}
\begin{aligned}
    h_t &= \overline{\textbf{\textit{A}}} h_{t-1} + \overline{\textbf{\textit{B}}} x_t \\ 
    y_t &= \textbf{\textit{C}} h_t + \textbf{\textit{D}} x_t,
\end{aligned}  
\end{equation}
where $\textbf{\textit{C}}$ and $\textbf{\textit{D}}$ represent projection parameters mapping the latent state and input to the output space. $\overline{\textbf{\textit{A}}}$ and $\overline{\textbf{\textit{B}}}$ are derived through the zero-order hold (ZOH) discretization method with a sampling step size $\Delta$. Specifically:
\begin{equation}
\begin{aligned}
    \overline{\textbf{\textit{A}}} &=  \operatorname{exp}(\Delta\textbf{\textit{A}}) \\
    \overline{\textbf{\textit{B}}} &= (\Delta\textbf{\textit{A}})^{-1}(\operatorname{exp}(\Delta\textbf{\textit{A}}) - \textbf{\textit{I}}) \cdot \Delta \textbf{\textit{B}}.
\end{aligned}  
\end{equation}

This discretization strategy ensures both numerical stability and computational efficiency. However, the parameters $\textbf{\textit{A}}$, $\textbf{\textit{B}}$, $\textbf{\textit{C}}$, $\Delta$ are fixed across all time steps, adhering to the Linear Time Invariant (LTI) property, which restricts the model's ability to accommodate dynamic variations in input sequences. To address this limitation, Mamba introduces a content-aware reasoning framework by modeling $\textbf{\textit{B}}$, $\textbf{\textit{C}}$, $\Delta$ as functions of the input. This transformation results in a selective SSM capable of adapting to input-dependent variations.

We feed the sequence of hypergraph features $\{\textbf{\textit{X}}'(t_i)\}$ into several Mamba blocks in reverse temporal order to capture the underlying propagation dynamics. Moreover, by utilizing convolution kernel $\overline{\textbf{\textit{K}}} = (\textbf{\textit{C}}\overline{\textbf{\textit{B}}}, \textbf{\textit{C}}\overline{\textbf{\textit{AB}}}, ..., \textbf{\textit{C}}\overline{\textbf{\textit{A}}}^k\overline{\textbf{\textit{B}}}, ...)$, we achieve the sequence-to-sequence transformation through $y = \textbf{\textit{X}}' * \overline{\textbf{\textit{K}}}$ with linear time complexity.

\paragraph{Graph-aware State Update Mechanism.}
While Mamba enables input-dependent learning with linear time complexity, directly feeding hypergraph snapshots as individual sequence elements into the model risks losing essential topological information. To address this, we propose a graph-aware state update mechanism allows for the update of node features by enabling node communication, where the state of each node is determined collaboratively by both the sequence input and the states of its neighbors, as expressed by:
\begin{equation}
h_t = \overline{\textbf{\textit{A}}} h_{t-1} + \overline{\textbf{\textit{B}}} x_t + h_{\mathcal{N}}.
\end{equation}

Similar to the convolutional process on hypergraphs, the neighbor state $h_{\mathcal{N}}$ is updated in two steps: node-to-hyperedge and hyperedge-to-node.

\textbf{(1) Node-to-Hyperedge} Acting as mediators between nodes, hyperedges first collect the averaged state information from its connected nodes:
\begin{equation}
    h_{edge} = \textbf{\textit{H}}^T\textbf{\textit{D}}_V^{-1}h_{t-1}.
\end{equation}

\textbf{(2) Hyperedge-to-Node} In this process, the state information is aggregated back to the nodes from the hyperedges:
\begin{equation}
    h_{\mathcal{N}} = \textbf{\textit{H}} \textbf{\textit{D}}_E^{-1}h_{edge}.
\end{equation}

Through aggregating the state information of neighboring nodes, the SSM acquires the capability to perceive the underlying graph structure. However, in the context of rumor propagation, distinct groups of nodes contribute differently to the spread of information. Therefore, hyperedges should be weighted based on their respective states. Given that attention mechanisms introduce additional computational complexity, we utilize a MLP block with an activation function and sigmoid to learn the weights of each hyperedge:
\begin{equation}
    \mathbf{\Omega}_e = \operatorname{sigmoid}(\sigma(\operatorname{MLP}(h_{t-1}))).
\end{equation}

Consequently, the state update process is summarized as:
\begin{equation}
    h_t = \overline{\textbf{\textit{A}}} h_{t-1} + \overline{\textbf{\textit{B}}} x_t + \textbf{\textit{H}} \textbf{\textit{D}}_E^{-1}\mathbf{\Omega}\textbf{\textit{H}}^T\textbf{\textit{D}}_V^{-1}h_{t-1}.
\end{equation}

The proposed graph-aware state update mechanism ensures that the node feature updates are informed by both the sequential data and the structural context of the hypergraph. Moreover, the adoption of hyperedge weights enables the model to dynamically adjust the influence of different hyperedges based on the propagation dynamics, providing a more adaptive and context-sensitive approach to learning rumor propagation patterns.

\subsection{Optimization and Training}
For the output sequence $y$, we focus on the last element, i.e., $y_1$, for source detection task. To alleviate the class imbalance resulting from the substantial disparity between the sample numbers of source and non-source nodes, we introduce a parameter $\xi$ to mitigate the prediction bias:
\begin{equation}
\xi = \frac{|s|}{n-|s|},
\end{equation}
where $n$ denotes the number of nodes, and $|s|$ represents the number of sources. This harmonic parameter equalizes the contribution of all samples. Consequently, the optimization objective is to minimize the discrepancy between the predicted set of sources and the actual sources, formalized as:
\begin{equation}
\min_GLoss = \sum_{v_{i} \in s} \mathcal{L}_{i}+\xi \sum_{v_{j} \in(V-s)} \mathcal{L}_{j}+\lambda\|w\|_{2},
\label{loss}
\end{equation}
where $\mathcal{L}$ denotes the cross-entropy loss, and for a sample $x$ with corresponding label $z$, $\mathcal{L}(x, z) = - \log(x) \times z$. The final term represents the $L_2$ regularization of the weight vector $\textbf{\textit{W}}$, with $\lambda$ controlling the regularization strength.

\begin{table*}[!h]
\centering
\resizebox{0.95\textwidth}{!}{%
\begin{tabular}{l|ccc|ccc|ccc|ccc}
\hline
\multirow{2}{*}{\textbf{Methods}} & \multicolumn{3}{c|}{\textbf{Zoo}}  & \multicolumn{3}{c|}{\textbf{House}}      & \multicolumn{3}{c|}{\textbf{NTU2012}}  & \multicolumn{3}{c}{\textbf{Mushroom}}  \\ \cline{2-13}
 & \textbf{ACC} & \textbf{F-Score} & \textbf{AUC} &  \textbf{ACC} & \textbf{F-Score} & \textbf{AUC} & \textbf{ACC} & \textbf{F-Score} & \textbf{AUC} & \textbf{ACC} & \textbf{F-Score} & \textbf{AUC} \\ \hline
LPSI & 0.802 & 0.345 & 0.779 & 0.812 & 0.347 & 0.807 & 0.787 & 0.321 & 0.821 & 0.815 & 0.329 & 0.775 \\
EPA & 0.789 & 0.351 & 0.785 & 0.814 & 0.340 & 0.811 & 0.794 & 0.312 & 0.820 & 0.817 & 0.326 & 0.774 \\
GCNSI & 0.823 & 0.372 & 0.810 & 0.828 & 0.329 & 0.811 & 0.817 & 0.306 & 0.825 & 0.818 & 0.362 & 0.811 \\
SIGN & 0.821 & 0.442 & 0.819 & 0.826 & 0.426 & 0.814 & 0.803 & 0.398 & 0.813 & 0.814 & 0.415 & 0.824 \\
GCSSI & 0.811 & 0.415 & 0.807 & 0.826 & 0.412 & 0.814 & 0.805 & 0.389 & 0.799 & 0.805 & 0.389 & 0.816 \\
ResGCN & 0.819 & 0.503 & 0.816 & 0.825 & 0.467 & 0.811 & 0.807 & 0.442 & 0.819 & 0.829 & 0.458 & 0.817 \\
IVGD & 0.847 & 0.589 & 0.835 & 0.824 & 0.529 & 0.836 & 0.859 & 0.532 & 0.856 & 0.851 & 0.526 & 0.864 \\
SL-VAE & 0.851 & 0.602 & 0.830 & 0.842 & 0.531 & 0.824 & 0.846 & 0.529 & 0.839 & 0.864 & 0.547 & 0.867 \\
GIN-SD & 0.861 & 0.694 & 0.834 & 0.859 & 0.624 & 0.854 & 0.860 & 0.634 & 0.841 & 0.874 & 0.642 & 0.884 \\
Ours & \textbf{0.915} & \textbf{0.797} & \textbf{0.920} & \textbf{0.938} & \textbf{0.836} & \textbf{0.941} & \textbf{0.947} & \textbf{0.869} & \textbf{0.944} & \textbf{0.929} & \textbf{0.827} & \textbf{0.930} \\ \hline

\multirow{2}{*}{\textbf{Methods}} & \multicolumn{3}{c|}{\textbf{ModelNet40}}  & \multicolumn{3}{c|}{\textbf{20News}}      & \multicolumn{3}{c|}{\textbf{PubMed}}  & \multicolumn{3}{c}{\textbf{Walmart}}  \\ \cline{2-13}
 & \textbf{ACC} & \textbf{F-Score} & \textbf{AUC} &  \textbf{ACC} & \textbf{F-Score} & \textbf{AUC} & \textbf{ACC} & \textbf{F-Score} & \textbf{AUC} & \textbf{ACC} & \textbf{F-Score} & \textbf{AUC} \\ \hline
LPSI & 0.784 & 0.254 & 0.771 & 0.745 & 0.273 & 0.741 & 0.759 & 0.215 & 0.764 & 0.718 & 0.112 & 0.718 \\
EPA & 0.791 & 0.258 & 0.797 & 0.771 & 0.284 & 0.767 & 0.774 & 0.205 & 0.794 & 0.715 & 0.108 & 0.730 \\
GCNSI & 0.814 & 0.278 & 0.816 & 0.805 & 0.237 & 0.801 & 0.798 & 0.202 & 0.791 & 0.808 & 0.174 & 0.810 \\
SIGN & 0.824 & 0.442 & 0.821 & 0.814 & 0.364 & 0.806 & 0.794 & 0.254 & 0.791 & 0.816 & 0.234 & 0.815 \\
GCSSI & 0.811 & 0.384 & 0.795 & 0.817 & 0.379 & 0.805 & 0.814 & 0.203 & 0.807 & 0.811 & 0.206 & 0.806 \\
ResGCN & 0.820 & 0.497 & 0.819 & 0.821 & 0.440 & 0.807 & 0.816 & 0.279 & 0.823 & 0.824 & 0.251 & 0.823 \\
IVGD & 0.820 & 0.629 & 0.827 & 0.815 & 0.557 & 0.823 & 0.819 & 0.563 & 0.817 & 0.810 & 0.547 & 0.823 \\
SL-VAE & 0.842 & 0.618 & 0.834 & 0.829 & 0.587 & 0.822 & 0.824 & 0.557 & 0.824 & 0.824 & 0.525 & 0.830 \\
GIN-SD & 0.836 & 0.612 & 0.824 & 0.831 & 0.584 & 0.819 & 0.822 & 0.587 & 0.828 & 0.831 & 0.536 & 0.825 \\
Ours & \textbf{0.921} & \textbf{0.814} & \textbf{0.926} & \textbf{0.908} & \textbf{0.792} & \textbf{0.918} & \textbf{0.905} & \textbf{0.784} & \textbf{0.897} & \textbf{0.879} & \textbf{0.715} & \textbf{0.865} \\ \hline
\end{tabular}%
}
\caption{The performance of source detection across all datasets for each method, with the best results being highlighted in bold.}
\label{overallperformance1}
\vspace{-1mm}
\end{table*}

\section{Experiments}
\subsection{Experimental Settings}
\paragraph{Datasets.}
Eight datasets with varying scales and characteristics are used to evaluate the performance of our SourceDetMamba. Considered datasets includes Zoo \cite{asuncion2007uci}, House \cite{chodrow2021generative}, NTU2012 \cite{chen2003visual}, Mushroom \cite{asuncion2007uci}, ModelNet40 \cite{wu20153d}, 20News \cite{asuncion2007uci}, PubMed \cite{yadati2019hypergcn} and Walmart \cite{amburg2020clustering}.

\paragraph{Baselines.}
Different categories of methods are selected as baselines, including source centrality-based methods such as LPSI \cite{wang2017multiple} and EPA \cite{ali2019epa}, user state-based methods like GCNSI \cite{dong2019multiple}, SIGN \cite{li2021propagation}, GCSSI \cite{dong2022wavefront} and ResGCN \cite{shah2020finding}, and propagation information-integrated methods such as IVGD \cite{wang2022invertible}, SL-VAE \cite{ling2022source}, and GIN-SD \cite{cheng2024gin}.

\paragraph{Implementation.}
Given the independence of users on social platforms and the short-term nature of rumor propagation, we simulate the rumor spread using a heterogeneous independent cascade model. The propagation process is triggered by randomly selecting 5\% of the nodes as sources. For low-order interactions, the propagation rate  $p_i$ of each node follows a uniform distribution $U(0, 0.5)$. In the high-order interaction phase, the state transition probability $p_{\Delta}$ for each node in a hyperedge is proportional to the number of informed nodes, formulated as $p_{\Delta}$ = $0.3\cdot(|e \cap v^+|/|e|), v^+ \in G_{t_i}^+$. The sequential snapshots are captured when 10\%, 20\%, and 30\% of the nodes in the network are informed, with the resulting dataset is split into training and testing sets in an 8: 2 ratio. For the training process, the model's learning rate and weight decay are set to \( 10^{-3} \) and \( 10^{-5} \), respectively. In the state space model, each input in the sequence undergoes a state update via a two-layer stack, with state size equals to 128. Hyperedge weights are calculated through two linear layers, each followed by an activation function, and the state is mapped to a single scalar value using a sigmoid function. The hypergraph snapshots are first subjected to clique expansion before being applied to the baseline methods. All experiments are conducted on a workstation equipped with four NVIDIA RTX 3090Ti GPUs.

\paragraph{Metrics.}

ACC (Accuracy), F-Score, and AUC (Area Under the Curve) are selected as evaluation metrics to assess the performance of different methods. ACC measures the proportion of correctly classified samples. F-Score comprises \textit{Precision} and \textit{Recall}, where \textit{Precision} = $|\hat{s} \cap s| / |\hat{s}|$ represents the proportion of true sources in the detected source set, and \textit{Recall} = $|\hat{s} \cap s| / |s|$ indicates the proportion of true sources detected. The AUC evaluates the model's ability to differentiate between source and non-source nodes across varying thresholds. These metrics provide a comprehensive evaluation of the performance and robustness of source detection methods.

\subsection{Performance Analysis}
\paragraph{Comparison with State-of-the-art Methods.}
Table \ref{overallperformance1} presents the comparative results of our proposed SourceDetMamba against baseline methods, revealing several key insights. First, all methods achieve higher ACC than F-Score, primarily due to the imbalance between the large number of non-source nodes and the relatively small number of source nodes. As ACC measures the overall proportion of correctly classified samples, this disparity highlights the impact of class imbalance, particularly for methods such as LPSI and EPA based on source centrality, and machine learning approaches like GCNSI, SIGN, and GCSSI. Second, methods relying solely on source centrality demonstrate lower F-Score due to the inherent randomness of the propagation process, which limits the effectiveness of topology-only approaches. Machine learning-based methods that incorporate user features perform better, and methods that model the propagation process, such as IVGD and SL-VAE, or integrate propagation features directly, like GIN-SD, achieve even greater performance improvements. Notably, SourceDetMamba outperforms all baselines, delivering 10\%-23\% higher accuracy than propagation-aware methods, 25\%-35\% higher than user state-based methods, and even doubling the performance of source centrality-based approaches. These significant improvements stem from SourceDetMamba's ability to directly learn propagation dynamics through Mamba blocks and its graph-aware state update mechanism, which seamlessly incorporates hypergraph topology into the learning process.

\begin{figure}[!htbp]
\vspace{-2mm}
  \centering
  \includegraphics[width=\linewidth]{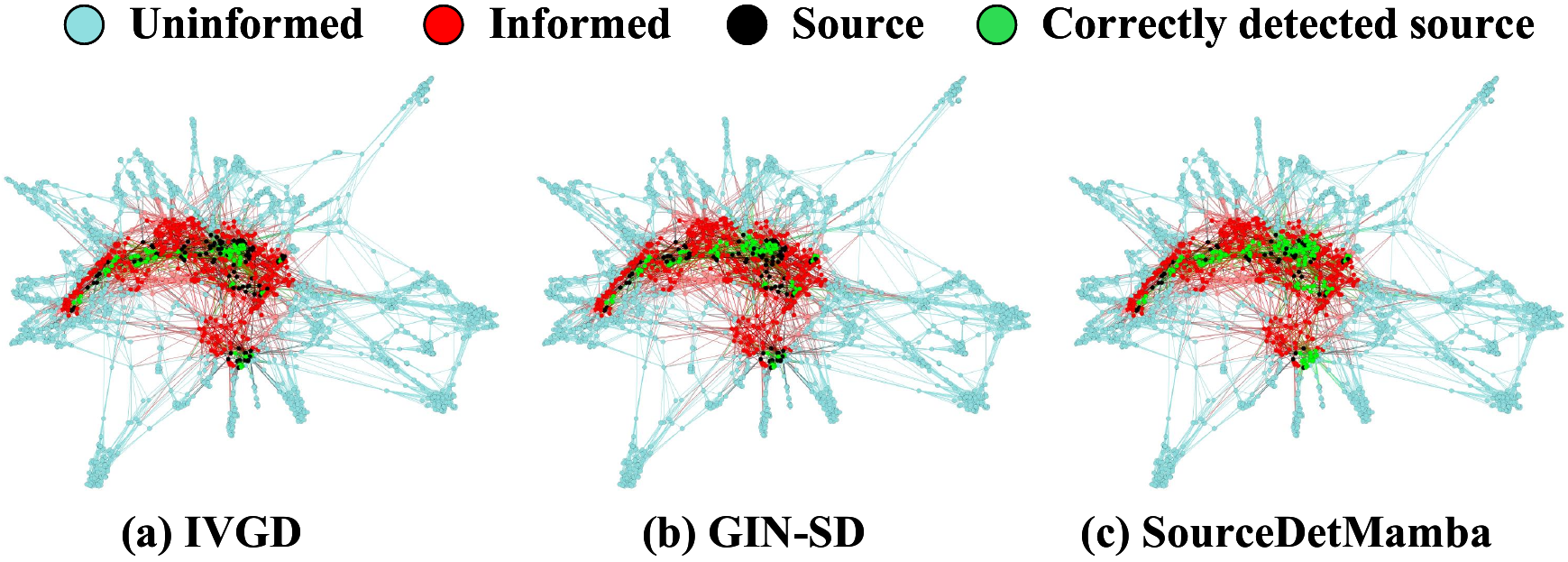}
  \vspace{-4mm}
  \caption{Visualization of source detection results on NTU2012.}
  \label{visualize}
  \vspace{-4mm}
\end{figure}

\paragraph{Visualization.}
To provide an intuitive illustration of the source detection results, we visualize the outputs of SourceDetMamba alongside those of representative methods on the NTU2012 dataset  \cite{chen2003visual} in Fig. \ref{visualize}. As compared to baselines, our approach demonstrates better performance in identifying the sources.

\begin{figure}[!tbp]
  \centering
  \includegraphics[width=\linewidth]{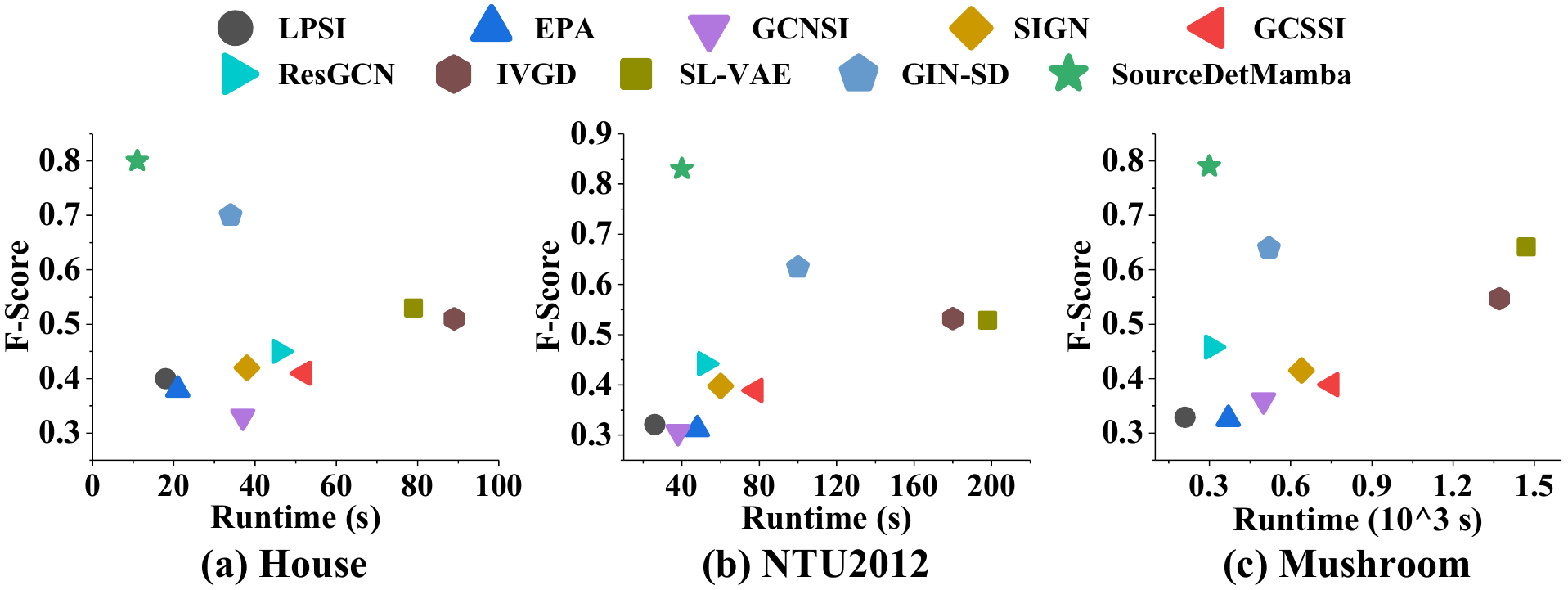}
  \vspace{-3mm}
  \caption{Comparison of runtime and F-Score across baseline methods, highlighting the balance achieved by SourceDetMamba.}
  \label{computational_efficiency}
  \vspace{-4mm}
\end{figure}

\paragraph{Computational Efficiency.}  
To evaluate computational efficiency, we compare the runtime and accuracy of the proposed method against baseline approaches, as shown in Fig. \ref{computational_efficiency}. The results reveal that methods not considering propagation dynamics achieve faster runtime but suffer from lower F-scores, indicating limited accuracy. Conversely, methods incorporating propagation dynamics demonstrate improved accuracy at the expense of significantly higher computational complexity. In contrast, our proposed method achieves a better balance, excelling in both accuracy and runtime efficiency. These findings validate the effectiveness of SourceDetMamba in delivering SOTA performance with computational efficiency.

\subsection{Ablation Study and Other Analyses}
\paragraph{Effects of Sequential Snapshots and Positional Encoding.}
We first remove the sequential snapshots, retaining only the earliest one. The results in Table \ref{ablation} show a noticeable performance decline compared to the complete model, highlighting the importance of utilizing multiple snapshots to capture propagation patterns. Next, we exclude the positional encodings in w/o PEs, the observed performance degradation highlights the critical role that PEs play in capturing the structural context and enhancing the model's robustness.

\paragraph{Effects of Exiting Sequential Models.}
We employ transformation and attention mechanisms to facilitate reverse learning of propagation patterns in w/ Attention, the results demonstrate a noticeable decline compared to SourceDetMamba. When sequential features are directly fed into an LSTM, the performance is similar to that of the attention-based model. In the w/ Att \& LSTM, we first apply attention fusion to the features and then input them into the LSTM. However, the results reveal a decrease in performance compared to using attention or LSTM individually. The performance is mainly attributed to the challenge of existing sequential models in effectively learning input dependencies.

\begin{table}[!bp]
\centering
\resizebox{\linewidth}{!}{%
\begin{tabular}{l|cccc}
\hline
{\textbf{Methods}} & {\textbf{House}}  & {\textbf{NTU2012}}  & {\textbf{20News}}  & {\textbf{PubMed}} \\ \hline
w/ Single snapshot & 0.758 & 0.768 & 0.694 & 0.672 \\
w/o PEs & 0.822 & 0.854 & 0.780 & 0.769 \\ \hline
w/ Attention & 0.798 & 0.805 & 0.703 & 0.712 \\
w/ LSTM & 0.785 & 0.794 & 0.691 & 0.716 \\
w/ Att \& LSTM & 0.782 & 0.806 & 0.683 & 0.714 \\ \hline
w/o Graph & 0.790 & 0.804 & 0.735 & 0.742 \\
w/o Edge weights & 0.824 & 0.847 & 0.772 & 0.759 \\
SourceDetMamba & \textbf{0.836} & \textbf{0.869} & \textbf{0.792} & \textbf{0.784} \\ \hline
\end{tabular}%
}
\caption{Performance of different SourceDetMamba variants.}
\label{ablation}
\end{table}

\paragraph{Effects of the Graph-aware State Update Mechanism.}
By removing node communication from Mamba and relying solely on the temporal states of individual nodes for prediction, we observe a significant performance drop. This highlights the critical role of the graph-aware state update mechanism in sequential graph tasks. Additionally, when hyperedge weight computation is omitted, performance improves compared to the model without the graph-aware mechanism, yet it still falls short of SourceDetMamba. These results confirm the effectiveness and validity of integrating topological structure and differentiating hyperedges in our proposed model.

\begin{table}[!tbp]
\centering
\resizebox{\linewidth}{!}{%
\begin{tabular}{c|cc|cc|cc}
\hline
\multirow{2}{*}{\textbf{Models}} & \multicolumn{2}{c|}{\textbf{House}}  & \multicolumn{2}{c|}{\textbf{ModelNet40}}  & \multicolumn{2}{c}{\textbf{PubMed}} \\ \cline{2-7}
 & \textbf{ACC} & \textbf{F-Score} & \textbf{ACC} & \textbf{F-Score} & \textbf{ACC} & \textbf{F-Score}\\ \hline
SI & 0.925 & 0.825 & 0.917 & 0.816 & 0.910 & 0.774 \\
SIS & 0.926 & 0.804 & 0.918 & 0.794 & 0.915 & 0.754 \\
SIR & 0.921 & 0.794 & 0.912 & 0.785 & 0.921 & 0.742 \\
IC & 0.938 & 0.836 & 0.921 & 0.814 & 0.905 & 0.784 \\ \hline
\end{tabular}%
}
\caption{Performance of SourceDetMamba under different information diffusion models.}
\vspace{-3mm}
\label{models}
\end{table}

\paragraph{Impact of Information Diffusion Models.}
To address diverse scenarios, we employ various information diffusion models and summarize the performance of SourceDetMamba in Table \ref{models}. The results demonstrate that SourceDetMamba consistently performs well across all tested models. A slight decline is observed for the SIS and SIR models due to the possibility of sources recovery, which introduces ambiguity into the detection process. Nonetheless, the strong overall performance underscores the robustness and adaptability of SourceDetMamba, reinforcing its effectiveness across a range of information propagation dynamics.

\begin{figure}[!htbp]
\vspace{-1mm}
  \centering
  \includegraphics[width=\linewidth]{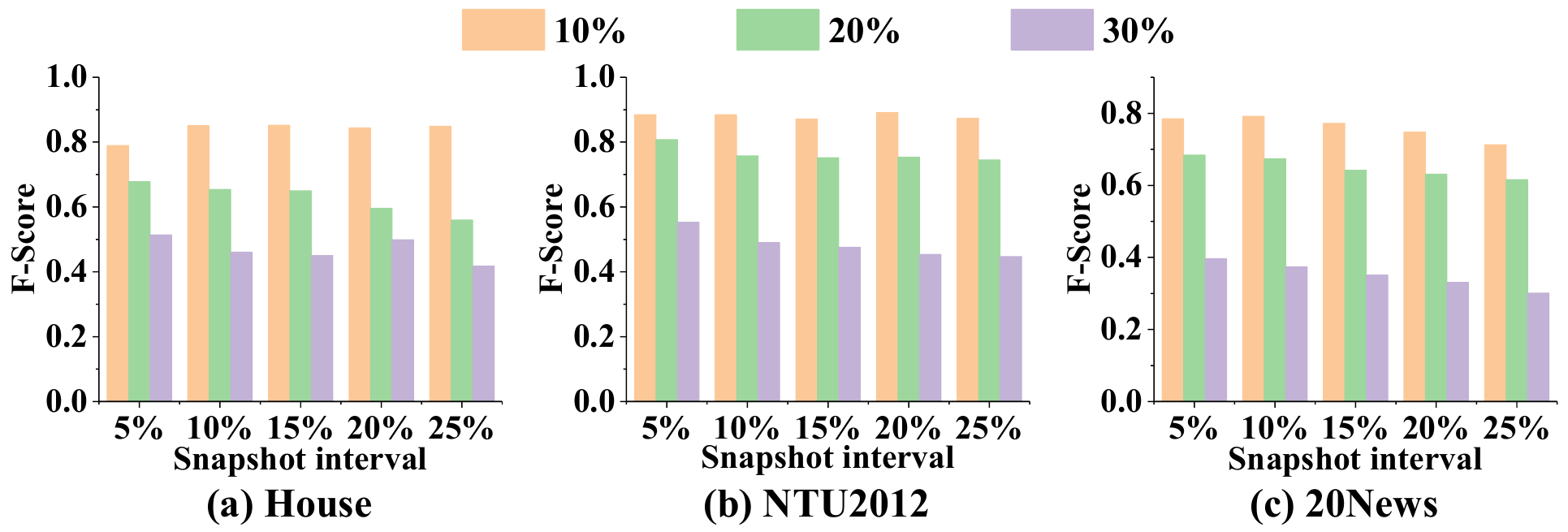}
  \caption{Performance of SourceDetMamba under different propagation scales and snapshot intervals.}
  \vspace{-5mm}
  \label{interval}
\end{figure}

\paragraph{Impact of Propagation Scale and Snapshot Intervals.}
We set the initial snapshot sizes to 10\%, 20\%, and 30\%, and vary the interval between subsequent snapshots from 5\% to 25\% with an increment of 5\%. The results of SourceDetMamba are shown in Fig. \ref{interval}. Firstly, across all datasets, the F-Score decreases as the propagation scale increases. This is primarily due to the necessity of identifying a larger number of samples, which leads to increased errors. Secondly, varying the snapshot intervals highlights that overly close snapshots may fail to capture sufficient propagation dynamics, while excessively distant snapshots may over-represent the influence of non-source nodes. Therefore, an appropriate snapshot interval is beneficial for effective source detection.

\section{Conclusion}
This paper presents a new perspective on rumor source detection by focusing on effective learning in sequential hypergraphs. The key idea involves reverse-order input of hypergraph snapshots into the Mamba state space model to learn propagation patterns, followed by the proposal of a graph-aware state update mechanism, which integrates topological information with temporal dependencies. Extensive experiments demonstrate that this approach significantly outperforms existing methods in sequential snapshot-based rumor source detection, providing a robust solution for sequential network learning. We hope that this work will inspire further research into topological-aware models in this field.

\section*{Acknowledgements}
This work was supported in part by the National Natural Science Foundation of China (Grant Nos. U22B2036, 62025602, 62471403, 62073263, 62472117), the Fundamental Research Funds for the Central Universities (Grant Nos. G2024WD0151, D5000240309), the Technological InnovationTeam of Shaanxi Province (Grant No. 2025RS-CXTD-009), the International Cooperation Project of Shaanxi Province (Grant No. 2025GH-YBXM-017), the Guangdong Basic and Applied Basic Research Foundation (Grant No. 2025A1515010157), the Science and Technology Projects in Guangzhou (Grant No. 2025A03J0137),  the National Foreign Experts Program (Grant No. G2023183019), and the Tencent Foundation and XPLORER PRIZE.

\bibliographystyle{named}
\bibliography{ijcai25}

\end{document}